 \documentclass[aps,prl,twocolumn]{revtex4-2}
 
\usepackage{mathbbol}
\usepackage{graphicx,color}
\usepackage{amsmath,amssymb,bm,amsthm,amstext}
\usepackage{mathtools}
\usepackage{dsfont}
\usepackage{simplewick}
\usepackage{braket}
\usepackage{qcircuit}
\usepackage{array}
\usepackage{soul}
\usepackage{dcolumn}
\usepackage[scr]{rsfso}

\begin{document}

\pagecolor{white}

\preprint{APS/123-QED}

\title{Microstructural Geometry Revealed by NMR Lineshape Analysis}

\thanks{Current address: Physics and Astronomy Department, UCLA}%

\author{Mohamad Niknam*}
\email{mniknam@physics.ucla.edu}
\affiliation{Department of Chemistry and Biochemistry, University of California Los Angeles, 607 Charles E. Young Drive East, Los Angeles, CA 90095-1059, USA}
\affiliation{Center for Quantum Science and Engineering, UCLA}
\author{Louis-S. Bouchard}
\email{lsbouchard@ucla.edu}
\affiliation{Department of Chemistry and Biochemistry, University of California Los Angeles, 607 Charles E. Young Drive East, Los Angeles, CA 90095-1059, USA}
\affiliation{Center for Quantum Science and Engineering, UCLA}

\date{\today}

\begin{abstract}
We introduce a technique for extracting microstructural geometry from NMR lineshape analysis in porous materials at angstrom-scale resolution with the use of weak magnetic field gradients. Diverging from the generally held view of FID signals undergoing simple exponential decay, we show that a detailed analysis of the line shape can unravel structural geometry on much smaller scales than previously thought. While the original q-space PFG NMR relies on strong magnetic field gradients in order to achieve high spatial resolution, our current approach reaches comparable or higher resolution using much weaker gradients. As a model system, we simulated gas diffusion for xenon confined within carbon nanotubes over a range of temperatures and nanotube diameters in order to unveil manifestations of confinement in the diffusion behavior. We report a multiscale scheme that couples the above MD simulations with the generalized Langevin equation to estimate the transport properties of interest for this problem, such as diffusivity coefficients and NMR lineshapes, using the Green-Kubo correlation function to correctly evaluate time-dependent diffusion. Our results highlight how NMR methodologies can be adapted as effective means towards structural investigation at very small scales when dealing with complicated geometries. This method is expected to find applications in materials science, catalysis, biomedicine and other areas.
\end{abstract}

\maketitle

\section{\label{sec:Intro}Introduction}

Extracting structural information at the angstrom scale directly from NMR lineshape analysis would mark a significant step forward in understanding molecular transport within confined spaces. Traditionally, pulsed-field gradient (PFG) NMR has been employed to investigate sample geometry, but this method necessitated multiple acquisitions of free induction decay (FID) signals at varying, often substantial, gradient field strengths, without directly analyzing the lineshape itself. FID signals were usually interpreted as simple exponential decays, which limited the structural insights that could be gained from the data. In contrast, our method emphasizes the analysis of NMR lineshapes under a weak gradient field, requiring only a single acquisition. This allows for the extraction of detailed geometric information from confined systems, offering a new way to explore microstructural properties without the need for repeated measurements.

This approach is especially beneficial in scenarios where confinement significantly impacts molecular dynamics, such as gases diffusing through carbon nanotubes (CNTs) or other porous materials. In these settings, restricted particle movement results in complex interactions with confining boundaries, which greatly affect diffusion behavior. The capability to investigate structural characteristics at the angstrom scale using weak magnetic field gradients would represent a substantial improvement over traditional q-space PFG NMR techniques, which generally depend on strong gradients to achieve high spatial resolution~\cite{Stejskal65,nature13,PRL15}.

A good understanding of molecular transport in confined geometries is vital not only in materials science and catalysis but also in biological and industrial contexts. In biological systems, diffusion plays a key role in processes like oxygen transport in tissues and drug delivery, or in the study of lung function using hyperpolarized gases. In industrial applications, diffusion is essential for operations such as catalysis, filtration, and separation within confined environments like porous materials~\cite{CALLAGHAN1990177,charati1998diffusion}. However, conventional methods often struggle to capture diffusion behavior in these complex systems, particularly when traditional Fickian diffusion models fail to account for confinement effects.

Herein we present a computational method that integrates molecular dynamics (MD) simulations with careful NMR lineshape analysis accounting for non-Markovian diffusion effects. This technique enables us to derive both diffusion coefficients and structural details from confined systems, such as CNTs, where the effects of confinement on molecular transport are particularly significant. By utilizing the generalized Langevin equation (GLE) and the Green-Kubo correlation function, we calculate time-dependent diffusion coefficients with high accuracy, offering deeper insights into the interactions between diffusing particles and the confining structures~\cite{zwanzig2001nonequilibrium,mcq76}.

We apply this method to investigate the diffusion of xenon gas in CNTs of different diameters, uncovering how the relationship between confinement and temperature influences diffusion behavior. The results indicate that in smaller CNTs, an increase in temperature results in slower diffusion due to stronger interactions with the tube walls. Conversely, in larger CNTs, we observe the anticipated increase in diffusion with rising temperature. The point at which this temperature dependence changes can act as an indicator of the geometry. These findings provide insights into the microstructural geometry of confined systems, with potential applications in PFG NMR, MRI, and other areas where molecular transport in confined geometries is crucial~\cite{niknam2023nuclear,JCP18}.

This new method can be viewed as a tool for examining molecular transport in confined environments, bridging the divide between MD simulations and experimental data. By showcasing the capability to investigate geometry at the angstrom scale using weak magnetic field gradients, we emphasize the potential of this approach to enhance the study of complex materials and catalysis, where conventional models often struggle.

\section{\label{sec:Theory} Theory}
\subsection{Generalized Langevin Equation (GLE)}

The non-Markovian dynamics of viscoelastic fluids are effectively described by the GLE, which provides a framework for incorporating memory effects into molecular transport:
\begin{equation}\label{eq:GLE}
 M\Dot{v} + \int_0^t \Gamma(t-\tau) v(\tau) \, \mathrm{d} \tau = \eta_f(t).
\end{equation}
Here, $M$ is the mass of the diffusing particle, $v(t) = \dot{x}(t)$ represents the particle's velocity, and $\eta_f(t)$ is a time-dependent stochastic force exhibiting colored noise behavior. Unlike in Markovian systems, where the force depends solely on the particle’s instantaneous velocity, in non-Markovian systems, the force depends on the entire velocity history. This introduces a memory function, $\Gamma(t)$, which accounts for the frictional forces experienced by the particle over time and is integrated over its past trajectory.

The memory kernel $\Gamma(t)$ is convolved with the particle’s velocity history to capture viscoelastic effects on the frictional forces. In general, the memory function depends on both wave-number and complex frequency, describing the system’s response to fluctuations across different temporal and spatial scales. This formalism suggests that the correlation functions of various dynamic properties—such as velocity autocorrelation, viscosity, and diffusion coefficients—decay according to the memory function. When written in terms of a normalized correlation function, $C(t)$, the GLE becomes the following integro-differential equation:
\begin{equation}
\dfrac{\mathrm{d}C(t)}{\mathrm{d}t} = -\int_0^t \Gamma(\tau) C(t-\tau) \, \mathrm{d} \tau,
\end{equation}
where $\Gamma(\tau)$ represents the memory kernel. This can be evaluated using projection operator techniques for the relevant variables~\cite{zwanzig2001nonequilibrium,mcq76,niknam2024nuclear}.

In this study, we focus on the time-correlation function of the pressure tensor as the primary tool for capturing memory effects. This is a critical quantity used to determine viscosity, which was chosen as the key parameter of interest due to its distinct temperature dependence in liquids versus gases. Moreover, viscosity can be directly accessed through MD simulations. Our focus on viscosity has proven highly effective, yielding accurate results that validate the robustness and efficacy of our approach.
 
\subsection{Generalized Stokes–Einstein Equation}

The diffusion of a spherical particle in a viscous fluid at low Reynolds numbers is classically described by the Stokes-Einstein equation:
\begin{equation}
D = \frac{k_B T}{6 \pi a \eta},
\end{equation}
where $D$ is the diffusion coefficient, $a$ is the radius of the diffusing particle, and $\eta$ is the zero-frequency shear viscosity of the surrounding fluid. This equation assumes a Markovian process, where memory effects are absent, and the mean-square displacement of diffusing particles increases linearly with time.

In more complex systems, particularly those exhibiting memory effects and non-Markovian dynamics, the Stokes-Einstein equation needs to be generalized to account for rheological effects. This generalization is typically performed by extending the equation into the Laplace domain, introducing a frequency dependence to the diffusion parameter. In the Laplace domain, the convolution integral in the GLE becomes a simple multiplication, allowing Eq.~(\ref{eq:GLE}) to be rewritten as:
\begin{equation}
\langle v(0) \tilde{v}(s) \rangle = \frac{k_B T}{M s + \tilde{\Gamma}(s)},
\end{equation}
where $s$ is the Laplace domain variable, and parameters with a tilde are understood to be analytically continued into the Laplace domain. In this equation, the term $M s$ accounts for inertial effects, which are typically negligible at low frequencies. The left-hand side can also be expressed in terms of the Laplace transform of the mean-square displacement, linking the memory function $\tilde{\Gamma}(s)$ directly to the fluid's viscosity.

This leads to a generalized Stokes-Einstein equation in the Laplace domain, expressed as~\cite{niknam2024nuclear,zwanzig1970hydrodynamic,mason1995optical,mason2000estimating,cordoba2012elimination}:
 \begin{equation}\label{eq:GSE}
\tilde{D}(s)=\frac{k_B T}{6\pi a s \tilde{\eta}(s)},
\end{equation} 
where $\tilde{D}(s)$ is the generalized diffusion coefficient, and $\tilde{\eta}(s)$ is the frequency-dependent viscosity. This generalized form has been experimentally validated using techniques such as diffusing-wave spectroscopy~\cite{mason1997diffusing}, illustrating its applicability to complex fluids and viscoelastic systems.

After obtaining the generalized diffusion function in the Laplace domain, an inverse Laplace transform is performed to recover the time-dependent diffusion coefficient, $D(t)$. This approach provides a robust tool for analyzing time-dependent transport phenomena in systems where memory effects play a significant role in governing diffusion behavior.

\subsection{Viscosity}

The autocorrelation function of the pressure tensor, commonly known as the Green-Kubo autocorrelation function, is used to derive the shear viscosity~\cite{mcq76,rapaport_2004,todd_daivis_2017,Brush62,RizkPRL2022}. The shear viscosity, $\eta$, in the long-time limit, is given by:
\begin{equation}
\eta = \lim_{t \to \infty} \eta_{GK}(t),
\label{eq:GKintegral}
\end{equation}
where $\eta_{GK}(t)$ is the Green-Kubo autocorrelation function, defined as:
\begin{equation}\label{eq:etaGK}
\eta_{GK}(t) = \frac{V}{3 k_B T} \int_0^{t} \sum_{\alpha < \beta} C_{\alpha \beta}(\tau) \, \mathrm{d} \tau,
\end{equation}
with $\alpha, \beta \in \{x, y, z\}$, $V$ as the system volume, and $T$ as the temperature. The function $C_{\alpha\beta}(\tau) = \left< p_{\alpha\beta}(\tau) p_{\alpha\beta}(0) \right>$ represents the autocorrelation of the off-diagonal elements of the pressure tensor. For instance, the $p_{xy}(t)$ component is given by:
\begin{equation*}
p_{xy}(t) = \frac{1}{V} \Biggl\{ \sum_j m_j v_{jx}(t) v_{jy}(t) + \frac{1}{2} \sum_{i \neq j} r_{ijx}(t) f_{ijy}(t) \Biggr\},
\end{equation*}
where $f_{ijy}$ is the $y$-component of the force between particles $i$ and $j$. The first term on the right-hand side denotes the kinetic contribution to the pressure tensor, while the second term accounts for the potential energy contribution. The other components of the pressure tensor, $p_{\alpha\beta}$, are defined analogously.

These correlation functions are computed numerically during the MD simulation, where they are evaluated as a function of time. The time-dependent autocorrelation functions are subsequently transformed into the Laplace domain, yielding the frequency-dependent viscosity, $\tilde{\eta}(s)$.

\begin{figure}[!htb]
\centering
\includegraphics[width=0.4\textwidth]{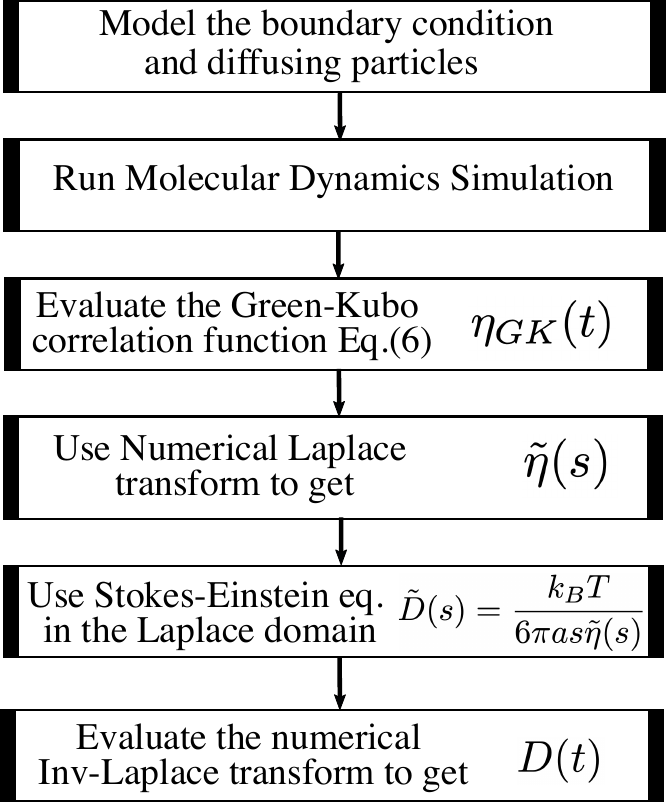}
\caption{Steps required to evaluate the generalized diffusion factor in the time domain. The Green-Kubo correlation function is first transformed into the Laplace domain to obtain the frequency-dependent viscosity function. The generalized Stokes-Einstein equation is then used to determine the frequency-dependent diffusion function, which can be transformed back into the time domain to derive the diffusion factor. All steps are performed analytically, and the code is publicly available at https://doi.org/10.5061/dryad.m905qfv7n.}
\label{fig:flowchart}
\end{figure}

Figure~\ref{fig:flowchart} provides a summary of the key steps involved in evaluating diffusion parameters for non-Markovian dynamics.

\section{\label{sec:CNTMD}Diffusion in CNTs}

The diffusion of gases within CNTs is an active area of research with important applications in energy storage, filtration, and environmental monitoring. CNTs, with diameters ranging from nanometers to angstroms, offer unique environments for gas diffusion, where molecular interactions with the inner walls of CNTs can substantially alter transport properties compared to bulk diffusion. Gases such as hydrogen, xenon (Xe), and methane are frequently studied in CNTs due to their relevance in applications like hydrogen storage, gas separation, and catalysis.

A key application of gas diffusion in CNTs is hydrogen storage. The adsorption properties of CNTs make them promising candidates for energy-efficient storage systems. Similarly, CNT membranes are used in gas separation and filtration technologies, offering selective diffusion of gas molecules, which is especially attractive for applications like CO\textsubscript{2} capture and air purification. In catalysis, the diffusion of reactants through CNT-based systems can significantly affect the efficiency and reaction rates, particularly within confined spaces.

In this study, we apply our methodology to explore the diffusion of gas particles in CNTs. This work demonstrates the capability of combining MD simulations and numerical analysis to model real systems and extract valuable transport parameters. The approach developed here is adaptable to other confined environments, providing a versatile tool for studying gas diffusion in nanoscale systems.

\begin{figure}[!htb]
\centering
\includegraphics[width=0.45\textwidth]{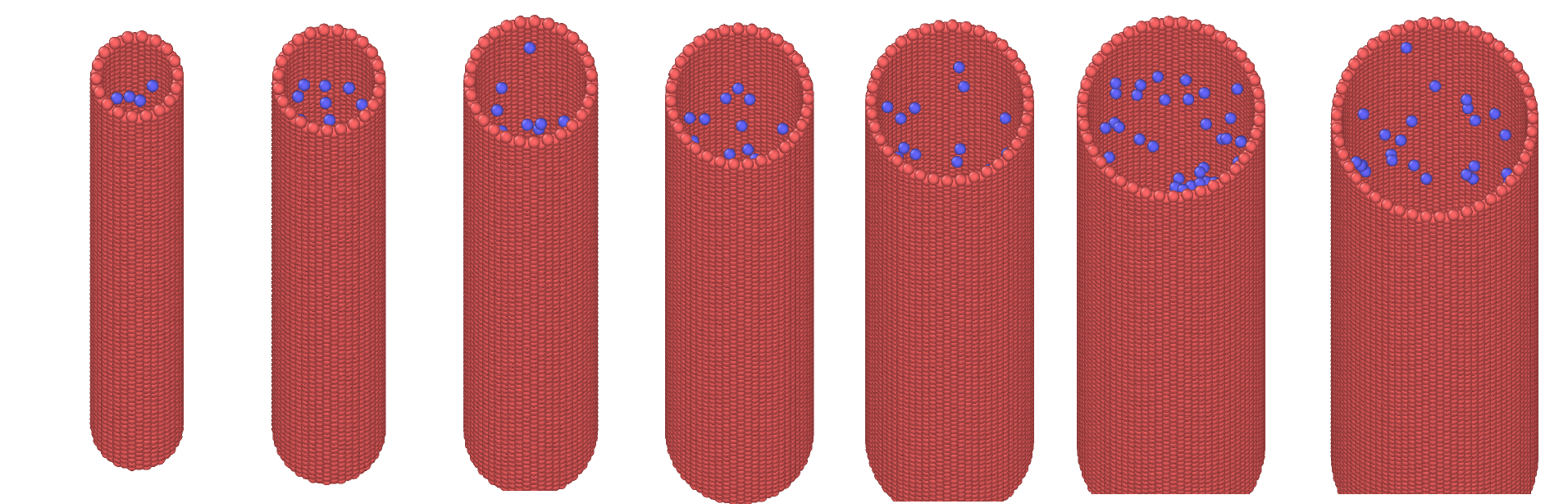}
\includegraphics[width=0.45\textwidth]{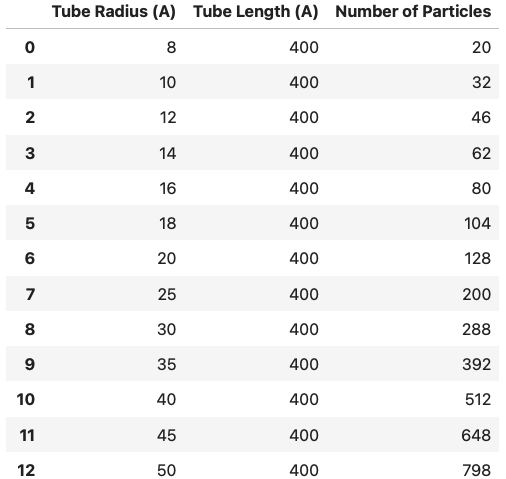}
\caption{Simulated structures of CNTs with diameters of 8, 10, 12, 14, 16, 18, and 20 Å, respectively, from left to right. Carbon atoms are shown in orange, while xenon (Xe) atoms are depicted as blue spheres. The number of Xe atoms is adjusted to maintain a consistent gas density across all CNT diameters.}
\label{fig:CNTs}
\end{figure}

As illustrated in Fig.~\ref{fig:flowchart}, the first step involves simulating the structure of CNTs and the diffusing gas particles. The CNT structure is generated by arranging carbon atoms in a cylindrical configuration, with each atom shifted by half a bond length every three rows. The code used for generating the CNT structure, provided in the supplementary material (\texttt{generate\_nanotube.lmp}), outputs the positions of the carbon atoms for CNTs with radii ranging from 8 to 50 Å, as shown in Fig.~\ref{fig:CNTs}.

The MD simulation evaluates the interactive forces between neighboring particles and their resulting motion. At each time step, the list of neighboring particles is updated, limiting the calculation of forces to relevant pairs for the next step. To improve efficiency, periodic boundary conditions are applied, allowing particles to exit the simulation box and re-enter from the opposite side. This approach enables the use of a finite number of particles while ensuring accurate transport calculations. Physical properties such as particle positions and momenta are sampled at predefined intervals, averaged, and recorded for analysis~\cite{FRENKEL200263,rapaport_2004}. 

The simulations were performed using the Large-scale Atomic/Molecular Massively Parallel Simulator (LAMMPS), an open-source MD simulation package that offers efficient and scalable implementations for large-scale simulations~\cite{LAMMPS}.

In this work, the primary output of the MD simulations is the Green-Kubo correlation function, as defined by Eq.~(\ref{eq:etaGK}), for Xe atoms diffusing inside CNTs. The Lennard-Jones (LJ) potential was used to model interactions between Xe particles:
\begin{equation}
U(r) = 4\epsilon\left[\left(\frac{\sigma}{r}\right)^{12} - \left(\frac{\sigma}{r}\right)^6\right],
\end{equation}
where for Xe-Xe interactions, $\epsilon = 1.77$ kJ/mol (potential well depth) and $\sigma = 4.1$ Å (the distance at which the potential energy becomes zero). Interactions between Xe atoms and the carbon atoms in the CNT walls were modeled with an LJ potential using $\epsilon = 0.71$ kJ/mol and $\sigma = 3.7$ Å. The carbon atoms were held fixed, and the length of the CNTs was set to 400 Å. The number of Xe atoms was adjusted according to the CNT radius to maintain a consistent gas density across all geometries.

Simulations were conducted over a temperature range of 240 K to 400 K, with 20 independent simulations performed at each temperature using different initial conditions. The Green-Kubo correlation function was sampled five times per simulation to improve statistical accuracy.

\begin{figure}[!htb]
\centering
\includegraphics[width=0.45\textwidth]{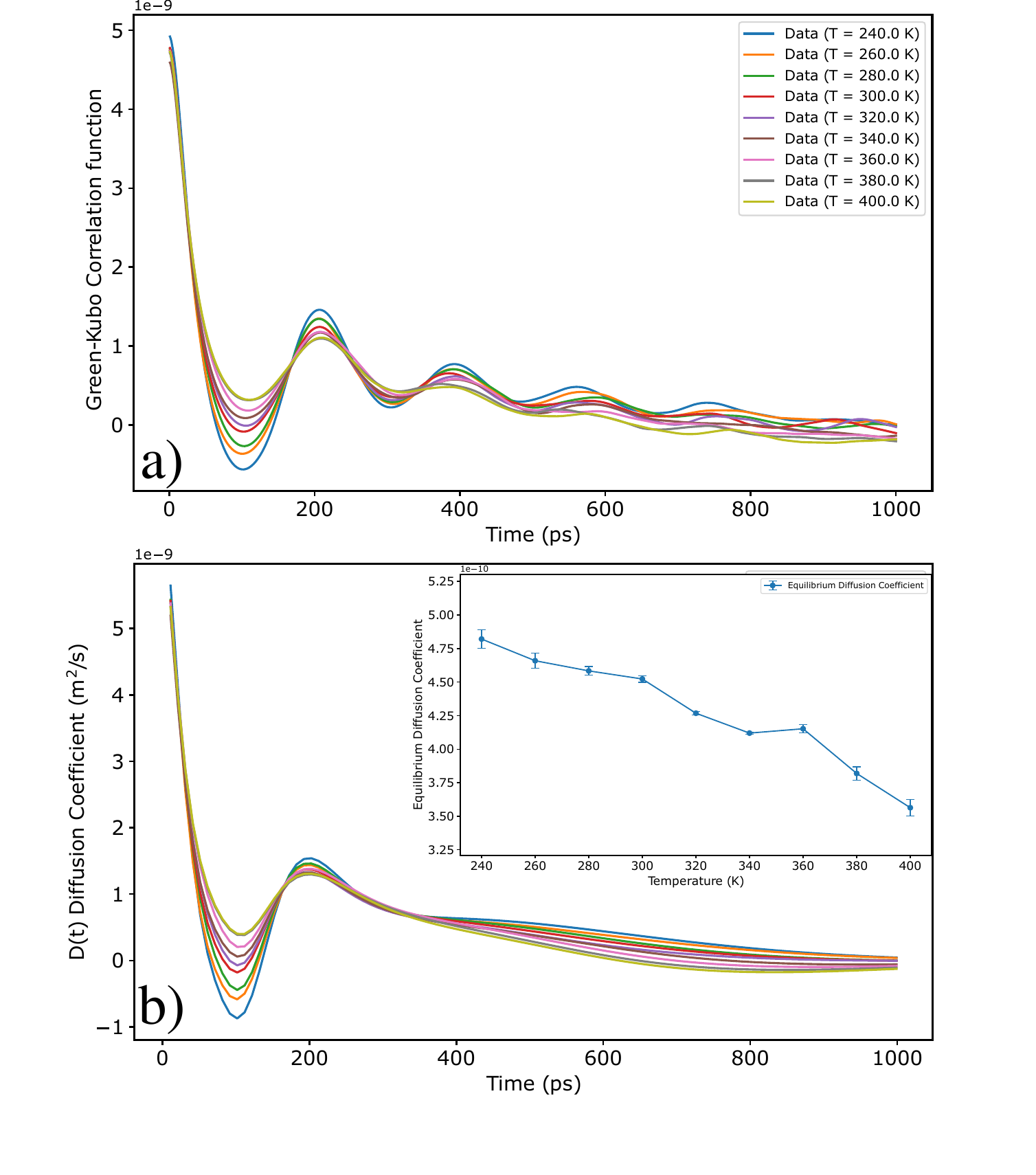}
\caption{(a) The Green-Kubo correlation function for Xe diffusion in a CNT with a radius of 16 Å at various temperatures. Each dataset represents an average over the dynamics of 80 particles, sampled 100 times. (b) The time-dependent diffusion factor obtained after applying the method described in Fig.~\ref{fig:flowchart}. The inset shows the equilibrium diffusion coefficients at various temperatures.}
\label{fig:16Adata}
\end{figure}

Figure~\ref{fig:16Adata}(a) shows the decay of the pressure tensor correlation function, also known as the Green-Kubo correlation function. The observed oscillations and long decay times in the correlation function indicate the presence of a memory function, reflecting the structured interactions between Xe atoms and the CNT walls. Previous studies have demonstrated that both the decay rate and oscillation frequency depend on temperature and the radius of the CNT~\cite{niknam2024nuclear}. This observation supports the notion that Xe atom dynamics in confined geometries significantly deviate from bulk behavior, with the memory function capturing the particle-wall interactions.

Figure~\ref{fig:16Adata}(b) presents the diffusion factor after applying the final three steps of the method outlined in Fig.~\ref{fig:flowchart}. The Laplace and inverse-Laplace transforms were performed using robust numerical methods, which are detailed in the supplementary material. The resulting time-dependent diffusion coefficient reaches an equilibrium value after a few nanoseconds, reflecting the steady-state diffusion behavior in the confined environment. The inset of Fig.~\ref{fig:16Adata}(b) shows the equilibrium diffusion coefficient at different temperatures.

\section{\label{sec:results} Diffusion Factor}

\begin{figure}[!htb]
\centering
\includegraphics[width=0.45\textwidth]{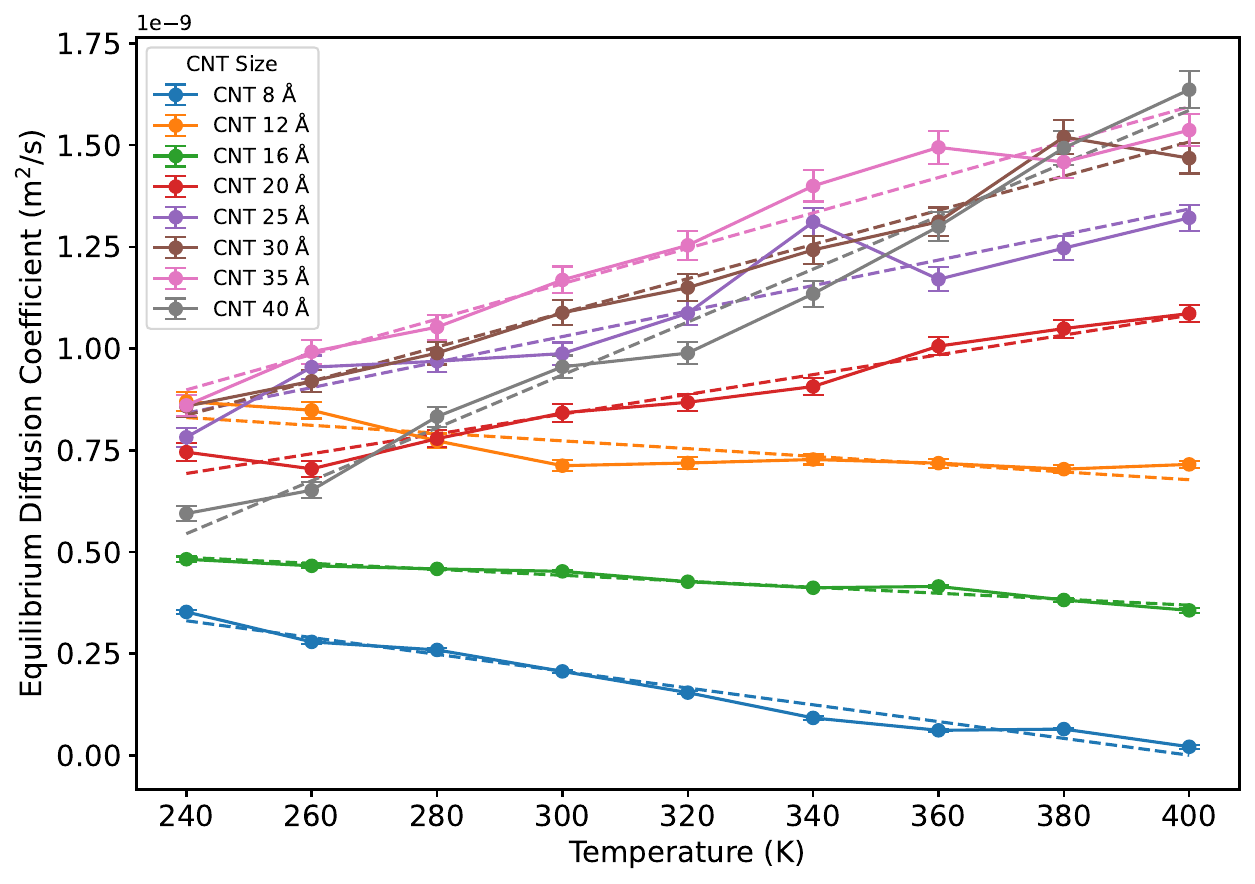}
\includegraphics[width=0.44\textwidth]{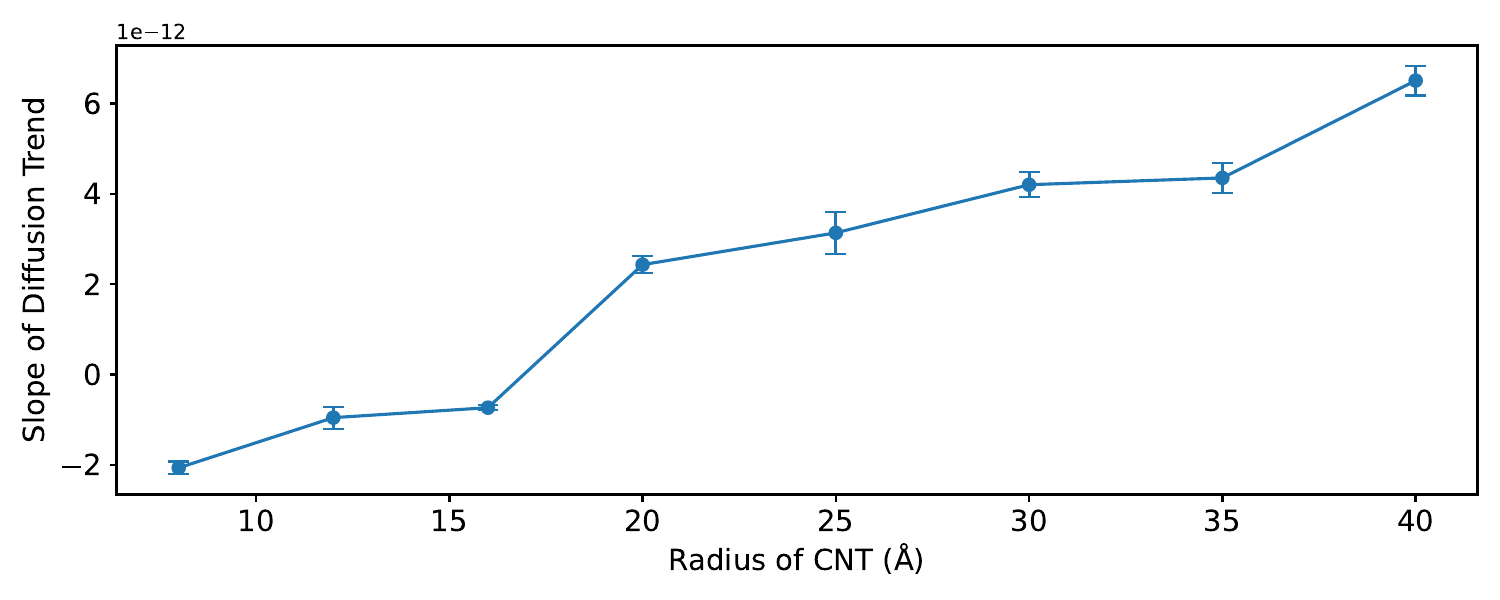}
\caption{(a) Diffusion coefficients calculated for various CNT sizes over a range of temperatures. For nanotubes with radii smaller than 20 Å, increasing temperature leads to slower diffusion, while for larger nanotubes, higher temperatures result in faster diffusion. (b) The slope of the linear relationship between temperature and diffusion coefficient serves as an indicator of nanotube size.}
\label{fig:diffc}
\end{figure}

We now examine the diffusion coefficients of Xe atoms across various CNT sizes and temperatures, as illustrated in Fig.~\ref{fig:diffc}. The results highlight significant differences in diffusion behavior due to the varying boundary conditions imposed by the size of the CNTs. Notably, for nanotubes with radii smaller than 20 Å, an increase in temperature leads to slower diffusion—this behavior contrasts with typical observations in bulk diffusion systems. In contrast, for nanotubes with radii larger than 20 Å, higher temperatures result in the expected behavior of faster diffusion.

This difference between smaller and larger CNTs can be explained by the degree of confinement experienced by the gas particles. In smaller nanotubes, the closeness of the CNT walls restricts the movement of Xe atoms, increasing collisions with the walls, which effectively reduces their mobility as the temperature rises. This phenomenon creates a unique transport regime where higher temperatures actually impede diffusion. Conversely, in larger nanotubes, the influence of wall interactions lessens, and the diffusion behavior aligns more closely with that of bulk systems, where increased thermal energy leads to faster particle motion.

Additionally, the diffusion coefficient shows a linear relationship with temperature. This relationship can be quantified by the slope of the diffusion coefficient's temperature dependence, which serves as a useful indicator of the nanotube's size, as shown in Fig.~\ref{fig:diffc}(b). This suggests that the slope of the temperature-diffusion relationship could be utilized as a diagnostic tool to infer the diameter of CNTs from diffusion measurements, establishing a direct link between diffusion behavior and the structural properties of the nanotube.

These findings offer valuable insights into the distinct diffusion regimes that arise in confined systems and emphasize the importance of accurately considering boundary conditions when modeling molecular transport in nanomaterials. The ability to extract information about the structural dimensions of CNTs through diffusion measurements has practical applications in fields such as nanotechnology, filtration, and catalysis.

\section{\label{sec:Applications}Applications}

One of the main questions that arises from this work is whether the changes observed in diffusion coefficients and their related structural features can be detected with sufficient sensitivity in experiments. PFG NMR is a commonly used technique for measuring diffusion coefficients. However, it relies on modulating the magnetization sinusoidally at a wavelength comparable to the spatial features under study.  When examining structures at smaller length scales, such as the nanoscale, PFG NMR usually requires strong magnetic field gradients to achieve the necessary spatial resolution. Traditional q-space PFG NMR experiments utilize gradients often exceeding several hundred mT/m to accurately spatially encode the positions of diffusing molecules~\cite{Stejskal65,callaghan1984pulsed}. This is required for resolving confined geometries, as the method relies on the interactions of molecules with walls and boundaries to significantly affect diffusion behavior on the length and time scales of the experiment.

The method introduced in this work greatly reduces the need for high-gradient fields. By analyzing the NMR lineshape with very weak magnetic field gradients (as low as 5-50 mT/m), we show the capability to investigate structural features at the angstrom scale. This represents a significant advancement, as weak gradients can still capture confinement effects and other geometric characteristics that were previously only accessible by applying much stronger gradients. In contrast, our method enables the extraction of detailed geometric information from confined environments, such as carbon nanotubes, without requiring extreme field gradients.

In a standard PFG NMR setup, a magnetic field gradient is applied to encode molecular positions based on the Larmor frequency at various spatial locations within the sample. The NMR echo sequence, $\frac{\pi}{2} - \tau - \pi - \tau$, with a constant gradient, $\mathbf{g}$, before and after the $\pi$-pulse, measures spin displacement by quantifying signal attenuation resulting from molecular motion. The attenuation function, $R(\tau)$, which describes the decay of the NMR signal as a function of molecular displacement, can be approximated for long diffusion times (longer than the memory function, typically on the order of 10 ns) as:
\begin{equation}
R(\tau) =  \left< \exp\left(- i \int_0^{\tau} \omega(t) \mathrm{d}t\right) \right> \approx \exp\left(- \gamma^2 g^2 D \tau^3\right),
\end{equation}
where $\omega(t) = \gamma g x(t)$ represents the frequency shift due to molecular displacement, $\gamma$ is the gyromagnetic ratio, and $g$ is the magnetic field gradient strength.

\begin{figure}[!htb]
\centering
\includegraphics[width=0.48\textwidth]{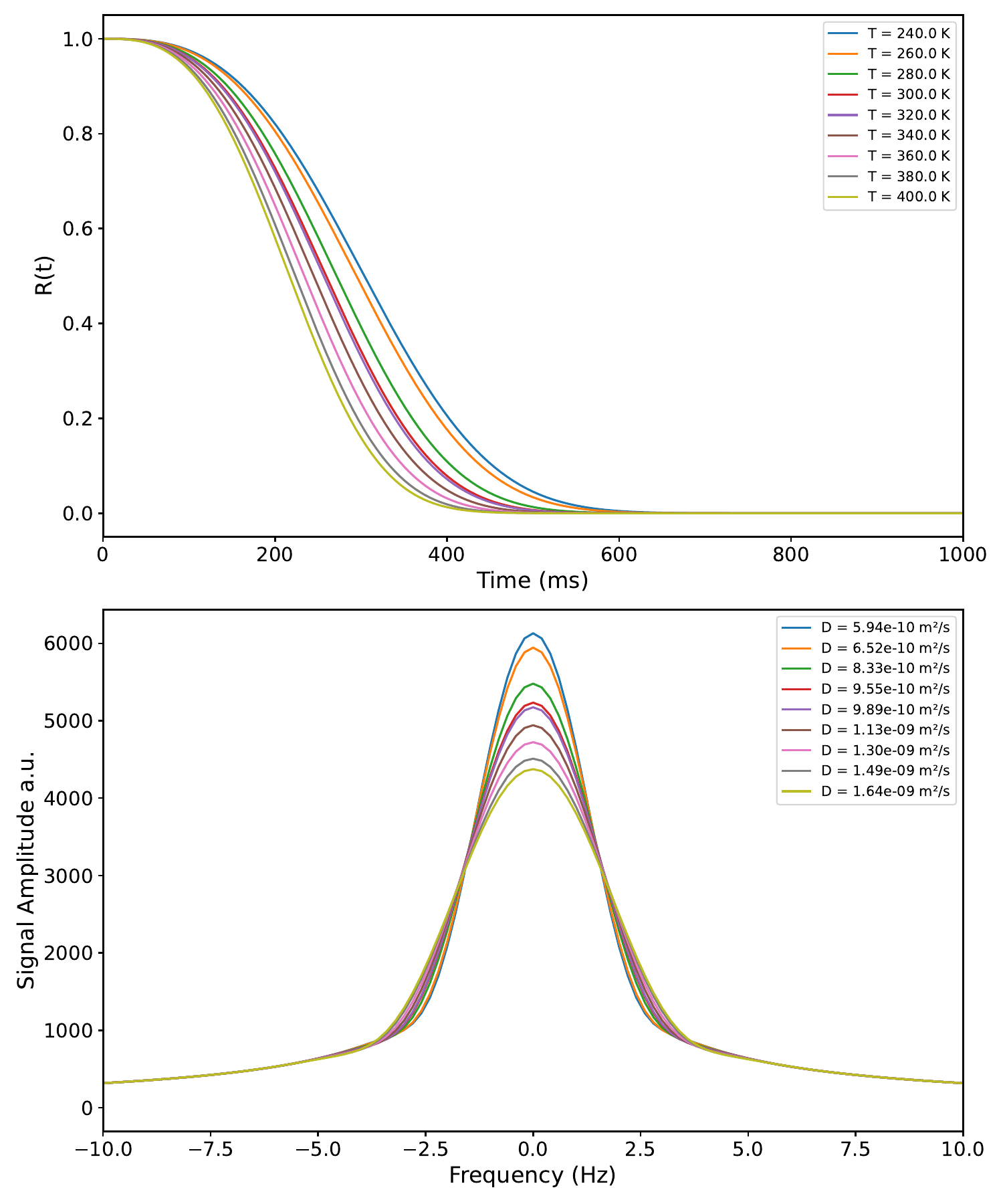}
\caption{(a) Attenuation function $R(\tau)$ for the PFG NMR signal of Xe atoms diffusing in a CNT with a 40 Å radius under a field gradient of 10 mT/m. (b) Fourier transformed lineshapes reveal a temperature-dependent change in linewidths, demonstrating that the variation in the diffusion coefficient due to the environment is detectable in the NMR signal.}
\label{fig:lineshape}
\end{figure}

Figure~\ref{fig:lineshape}(a) shows the attenuation function, $R(\tau)$, for Xe atoms diffusing in a CNT with a 40 Å radius under a relatively weak gradient field of 10 mT/m. These gradient values are typically used in medical imaging, usually falling between 5 and 50 mT/m. The diffusion coefficients for Xe particles are calculated over a range of temperatures (240 K to 400 K), illustrating how confinement influences the diffusion process. Figure~\ref{fig:lineshape}(b) presents the corresponding Fourier transformed NMR lineshapes, which display distinct changes in linewidths that depend on temperature. This variation in linewidths is directly related to changes in the diffusion coefficient, confirming that geometric effects in the diffusion environment can be detected through NMR lineshape analysis.

The key advantage here is that by examining the lineshape under weak gradient conditions, we can investigate structures on the angstrom scale—far exceeding the typical resolution of conventional PFG NMR. This approach removes the necessity for strong gradients and greatly expands the technique's applicability for studying confined systems and materials with nanoscale features.

These findings underscore the potential of PFG NMR as a sensitive method for identifying variations in molecular diffusion due to structural and environmental changes. The temperature-dependent changes in linewidths, as illustrated in Fig.~\ref{fig:lineshape}(b),  offer valuable insights into confinement effects, material properties, and the interaction strengths between diffusing gas molecules and their surrounding medium.

This approach can be used to investigate the structural characteristics of nanomaterials, porous systems, and confined geometries, where diffusion behavior is particularly influenced by boundary conditions and molecular interactions. It offers a powerful new method for analyzing molecular transport and material properties with exceptional resolution, eliminating the necessity for large gradient fields.

\section{Conclusion}

In this study, we introduced a numerical simulation technique to explore diffusion in complex environments where non-Markovian behavior is significant. By utilizing memory effects derived from MD simulations, which are usually used for calculating viscosity parameters, we integrated these insights with a model-free approach. Rather than making assumptions about the system’s underlying dynamics, we employed numerical Laplace transforms, the generalized Stokes-Einstein equation, and inverse Laplace transforms to accurately capture transport properties in confined geometries.

We showcased the flexibility of this method by calculating the diffusion coefficient of Xe atoms in CNTs with varying diameters. Our findings indicate that diffusion behavior is highly sensitive to the geometry of confinement. A notable difference from previous studies is that our analysis relies on the shape of the FID, offering deeper structural insights. For CNTs with radii smaller than 20 Å, increasing temperature leads to slower diffusion due to stronger interactions with the confining walls, while for larger CNTs, the diffusion coefficient rises with temperature, as expected in bulk-like systems. This connection between geometry and temperature highlights the potential of our method for extracting valuable structural information from confined systems.

A key advancement discussed in this work is the ability to explore structural features at the angstrom scale using weak magnetic field gradients. This marks a notable shift from traditional q-space PFG NMR techniques, which usually rely on very strong gradients to identify nanoscale features. By examining the NMR lineshape with weak gradients, we show that we can gather geometric information from a single measurement, removing the need for multiple tests with different gradient strengths. This opens up new avenues for investigating confined environments, ranging from nanomaterials to biological systems, with exceptional resolution.

These results highlight the potential of NMR techniques not just to examine diffusion properties but also to deduce the geometric characteristics of the confining environment through careful analysis of the FID lineshape. This approach sets the stage for innovative NMR diffusion experiments, with possible applications in MRI, materials science, and catalysis, where a deep understanding of the microstructure of complex systems is crucial. Expanding this method to other confined geometries, like porous materials or biological systems, could significantly enhance our understanding of molecular transport and confinement across a wide range of applications.

\section{Acknowledgments}
The computational and storage services used in this work are part of the Hoffman2 Shared Cluster, provided by the UCLA Institute for Digital Research and Education’s (IDRE) Research Technology Group. L.-S. B. acknowledges financial support from NSF award CHE-2002313.

\section{Data Availability Statement}

The data that support the findings of this study are available from the corresponding author upon reasonable request. 
Jupyter notebooks used in this study can be downloaded at:
\texttt{https://doi.org/10.5061/dryad.m905qfv7n}

\bibliographystyle{naturemag}
\bibliography{MD3.bib}

\end{document}